\documentclass[aps,twocolumn,groupedaddress,showpacs,prb]{revtex4}
\usepackage{graphicx}

\begin{document}

\title{Digital Darkfield Decompositions}
\author{P. Fraundorf}

\email[]{pfraundorf@umsl.edu}
\affiliation{Physics \& Astronomy/CME, U. Missouri-StL (63121), St. Louis, MO, USA}

\date{\today}

\begin{abstract}
Lattice imaging facilitates {\em digital} implementation of optical 
darkfield strategies that have for decades played a key role in the 
conventional electron microscopy of materials.  Applications to the 
microscopy of periodic structures are described in the context of 
recent developments in the mathematical harmonic analysis community, 
with hopes of inspiring further development along these lines.  
Applications here, which push present day limits using sharp-edged Fourier 
windows alone, include the location of weak periodicities in images 
by virtue of their spatial correlations, and the quantitative mapping 
of projected strain in a variety of nanostructures.
\end{abstract}
\pacs{03.30.+p,01.40.Gm,01.55.+b}
\maketitle







\affiliation{Physics \& Astronomy, U. Missouri-StL (63121), St. Louis, MO,
USA}




\section{Introduction}


The word wavelet is less familiar in microscopy than in other areas of applied math. However, microscopists have long been versed at interpreting images which trade spatial resolution for information on transverse momentum (e.g. on scattering direction, for example). As a result, microscopists (particularly those involved in transmission electron microscopy of crystals) have been developing skills connected to the use of combined space and frequency decompositions since well before the coining of the word wavelet in the 1980's \cite{Goupillaud84}. This is partly because microscopists do their decompositions with help from optical (rather than digital) computation, wherein their analysis hides behind the more familiar appellation: darkfield imaging (cf. Hirsch et al. \cite{Hirs11}). 

Here we discuss a few ways microscopists, and other periodicity analysts as well, might bring evolving digital tools of harmonic analysis to bear on problems traditionally addressed in analog fashion on live images using sharp-edged back focal-plane apertures and a variety of time-honored optical darkfield techniques.  In the process we seek to bring their challenges to the attention of the oft-times application-inspired harmonic analysis community \cite{Daubechies90, Chui96}.

\section{The darkfield image transform}

The elementary space plus frequency decomposition in signal processing is sometimes viewed optically as a windowed Fourier transform, in which the window and exponential kernel of the original transform combine to become the new kernel of the harmonic transform \cite{Boone98}. In this context, such ``windowlet'' transforms in 1D might be written as
\[
\psi \left[ {t,f_o ,\Delta f} \right] \equiv \Delta f\int\limits_{ - \infty }^{ + \infty } {s\left[ {t'} \right]h\left[ {\Delta f(t' - t),\frac{{f_o }}{{\Delta f}}} \right]dt'} 
\]
where s[t] is the input function, $t$ is a direct-space (or time) shift parameter, $f_o$ is a frequency shift parameter, and $\Delta f$ is a frequency-space scale parameter (reciprocal to the position/time-scale parameter $\Delta t$).  Note that we are using dimensioned parameters here, even though the kernel function $h[\xi ,\eta]$ is a function of dimensionless quantities.  If the kernel can be written as the product of a window function $w[\xi]$ and a modulation function $g[\xi ,\eta]=e^{-2\pi i \xi \eta}$, then the Fourier transform of the windowlet becomes simply
\[
\Psi \left[ {f,f_o ,\Delta f} \right] = S\left[ f \right]W\left[ {\frac{{f - f_o }}{{\Delta f}}} \right]
\]
where $S[f]$ and $W[\eta]$ are the Fourier transforms of $s[t]$ and $w[\xi]$,
respectively.

When the Fourier window is the simplest type of optical window to manufacture, namely a sharp-edged aperture of width $\Delta f$ centered at frequency $f_o$, then $W[\eta] = Rect[\eta] \equiv$ If$[{\vert}\eta{\vert}{<}\frac 12,1,0]$.  Hence the direct space window is $w[\xi]=\frac{\sin [\pi \xi]}{\pi \xi} \equiv$ sinc$[\xi]$.

\begin{figure}
\includegraphics{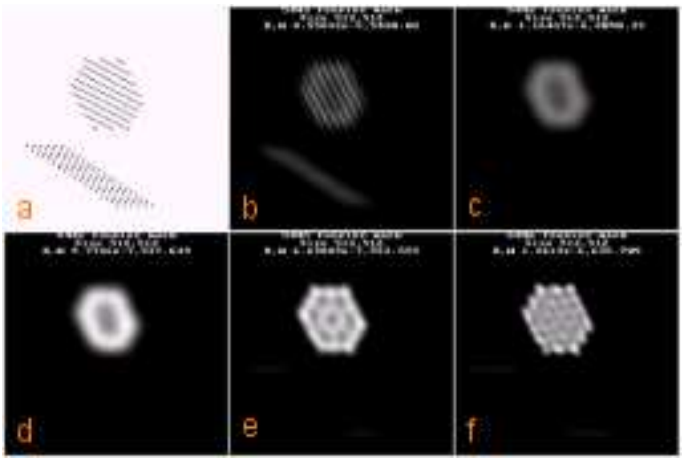} 
\caption{Amplitude images of a 2D spatially-delimited periodic array (e.g. a small crystal in projection). Inset a) is the starting brightfield dataset; b) is an image formed from a conjugate symmetric pair of diffracted beams and the DC peak; c) is the same as inset b without the DC peak -- note the doubled periodicity because of the doubled spacing between interfering tiles; d) is the amplitude of a complex darkfield formed using a rather small single tile; e) is the same as inset d using a larger tile in frequency space; f) is the same as inset d using an even larger tile in frequency space -- note the uncertainty principle in action in this last series of three.}
\label{Fig1}
\end{figure}

When ${\vert}f_o{\vert}{<}\frac{\Delta f}{2}$, microscopists refer to the map of intensity $\psi ^* \psi$ as a {\em brightfield image} (exclusively so when illumination is parallel) because it includes the unscattered beam.  Regions in the image without scatterers are therefore bright.  Conversely, when the image does not include the unscattered beam microscopists call the $\psi ^* \psi$ map a {\em darkfield image}.  The corresponding phase maps, by comparison, may contain very precise information on the lattice strain tensor in projection.  This transform lends itself to rapid calculation of a series of fast Fourier darkfield decompositions that bridge the gap in a variety of ways between direct and reciprocal versions of a given dataset.  Issues related to integrability, admissibility, and regularity are considered in more detail below.

Because this sinc (or Shannon/Littlewood-Paley) windowlet is the one with which darkfield microscopists have the most optical experience, it's uses will be the focus of this paper.  We expect, for work on atomic and lattice resolution digital images, that applications familiar to microscopists optically will benefit from the much more diverse set of transforms possible digitially.  One purpose of this paper is therefore to bring these applications to the attention of the signal processing community.

In particular, strategies for reciprocal space ``tiling'' so as to sample all of frequency space are an area of active interest in the signal processing community today (cf. Flesia et al. \cite{Flesia01}). However, the focus is often largely on compression techniques.  A very different set of considerations is at work in the microscopy community, particularly in the study of high resolution transmission electron microscope images of periodic lattices, and defects or interfaces associated therewith.  With recent developments in atomic-resolution digital microscopy allowing acquisition of near-gigabyte images by a variety of techniques, the opportunities for signal processing detective work on the atomic scale are quite rich.

Before we move on to the 2D transform, we should ask whether this choice of $w[\xi]$ meets the usual qualifications for wavelet decomposition.  Although it is a direct-reciprocal reversal of the ``short-time Fourier window'' decomposition oft mentioned in the literature on wavelets, this sinc-function window has zero mean and is square integrable.  It's transform also meets the admissibility condition (at least in the parallel illumination case) provided ${\vert}f_o{\vert} > \frac{\Delta f}{2}$, i.e. when the frequency window does not include the zero-frequency point or ``DC peak''.  This of course is consistent with the standard definition of ``darkfield image'' in microscopy, i.e. an image formed from waves which do not include the unscattered beam so that regions free of scatterers look dark.  The weak point of the decomposition, of course, is that it exhibits sinc-system convergence $\propto \frac{1}{t}$ as $t \to \infty$, at one end of the Harr to Sinc system continuum \cite{Vetterli95}.  Although the number of vanishing elements required depends heavily on the application \cite{Daubechies98}, this is a practical handicap for example when locating nanocrystal edges in otherwise noisy images.  Improved spatial convergence is hence one area where digital signal processing (for example by refining a decomposition to recognize lattices of discrete bumps in the 0.2 nanometer size range) might contribute downstream.  Even the utility of existing decompositions with good frequency resolution (e.g. brushlets \cite{MeyerCoifman97} and ridgelets \cite{Flesia01}) for these applications remains to be examined.

In the 2D case, using cartesian vectors and a common bandwidth $\Delta f$ for each direction, the windowlet transform becomes 

\[
\psi \left[ {{\bf x},{\bf f}_{\bf o} ,\Delta f} \right] \equiv \left( {\Delta f} \right)^2 \int\limits_{ - \infty }^{ + \infty } {s\left[ {{\bf x'}} \right]h\left[ {\Delta f({\bf x'} - {\bf x}),\frac{{{\bf f}_{\bf o} }}{{\Delta f}}} \right]d^2 {\bf x'}} 
\] If the kernel $h$ can further be written as the product of a window function $w\left[{\bf \xi} \right]$ and a modulation function $g\left[ {{\bf \xi },{\bf \eta }} \right] = e^{ - 2\pi i{\bf \xi } \bullet {\bf \eta }}$, the Fourier transform of the windowlet becomes

\[
\Psi \left[ {{\bf f},{\bf f}_{\bf o} ,\Delta f} \right] = S[{\bf f}]W\left[{\frac{{{\bf f} - {\bf f}_{\bf o} }}{{\Delta f}}} \right]
\]  Finally, a square Fourier window (for simplicity) looks like $W[\eta_x , \eta_y] =$ = Rect[$\eta_x$] Rect[$\eta_y$].  This results in a 2D direct-space sinc-window of the form $w\left[ {\xi _x ,\xi _y } \right] = \frac{{\sin \left[ {\pi \xi _x } \right]}}{{\pi \xi _x }}\frac{{\sin \left[ {\pi \xi _y } \right]}}{{\pi \xi _y }}$.  The shape of the Fourier window is of course arbitrary mathematically.  For optical applications, particularly in the electron microscopy case when the physical aperture size is in the 10 to 100 micron size range, circular rather than square apertures are more common (and easier to manufacture as well).

\section{Non-tiled Circular, Gaussian, and Bayesian Background-Subtracted Darkfields}

\begin{figure}
\includegraphics{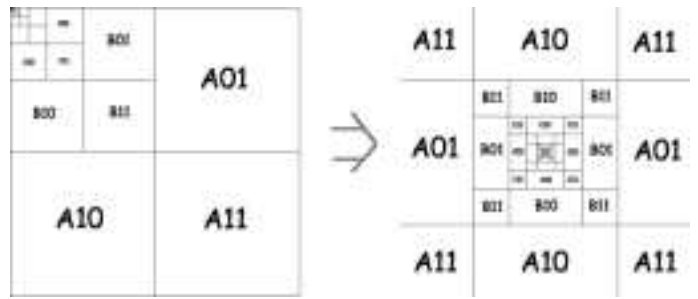} 
\caption{Left: (a) Traditional wavelet tiling of reciprocal space; Right: (b) Wavelet tiling rearranged into ``diffraction pattern'' format, with a centered DC peak.}
\label{Fig2}
\end{figure}

One feature of the sinc windowlets defined above, which may at first glance be unfamiliar to microscopists, is that they are generally complex. Fourier filtering of initially real images traditionally preserves their conjugate symmetry in reciprocal space, via selection of regions (e.g. to window) that are symmetric about the DC peak \cite{Fraundorf91}. The complex darkfield images described here do not. As a result, however, they reclaim something that is normally lost in Fourier filtering of images:  The continuous (fringe-free) illumination of scattering centers (e.g. of crystals in the field diffracting electrons into the selected Fourier aperture). This is because darkfield imaging with windows which may be asymmetric about the DC peak, of the sort described here, relegates periodicity fringes in the images largely into phase, rather than amplitude, variations in the image. By simply examining amplitude (or amplitude squared) across the images, these variations disappear just as they do in the microscope when image intensities (and not phases) are recorded in the imaging process. These advantages of asymmetric darkfield are illustrated in Fig. \ref{Fig1}. 


\begin{figure}
\includegraphics{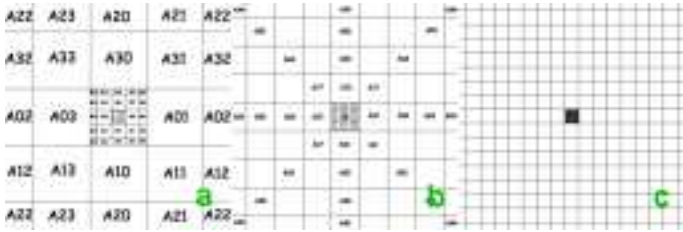} 
\caption{Linear decomposition heirachies with 4-fold symmetry: (a) $4 \times 4$, (b) $8 \times 8$, (c) $16 \times 16$.}
\label{Fig4}
\end{figure}

As long as tiling is not a requirement of the analysis process, for example when one simply wants to see where in the image the source of a peak in the power spectrum is located, any imaginable window function (not including the DC peak) may be used to form a darkfield image. Given the growing ability of microscopes to provide large digital images containing abundant crystal lattice fringe information, one can thus use these wavelets in the computer to do darkfield imaging just as in a transmission electron microscope, with the added advantage that one can be much more creative, and less limited by the physical challenges of aperture design, than is possible in the microscope. 

For example, microscope apertures usually have circular symmetry (at least by design).  In the case of a circular opening in a microscope's back-focal plane, $W[\eta, \phi] = Rect[\eta]$ in polar coordinates, and the direct-space window function becomes the well-known Airy distribution $w[{\xi, \phi}] = \frac{{J_1 [2\pi \xi ]}}{{2\xi }}$ expressed in terms of the Bessel function of the first kind $J_n$.  Although sharp-edged apertures are often easiest to manufacture (at least for electron optic applications), convergence in mathematical applications can be improved by the use of soft-edges (e.g. the Gaussian Fourier window functions associated with the Morlet wavelet).  Creative choice of the window function can do more than improve regularity.  For example, such functions can be ``tuned'' with Weiner filtering for the Bayesian removal of broad spectrum background in images \cite{Marks96, Fraundorf90}, or for the recognition of patterns of interest in images, such as edges.  Digital darkfield analysis, in that sense, provides yet another context for their continuing refinement.

In addition to use of different ``back-focal apertures'' of the form $W[{\bf \eta}]$, this strategy also facilitates a number of ``decoherence'' strategies, some of which (like hollow-cone darkfield) have been implemented optically, and others which have not. This is done by forming separate darkfield images in more than one chosen (typically local) range of frequencies, and then either combining or cross-correlating the intensities (square amplitudes) found therein. Thus annular darkfield images might be formed by adding intensities in patches forming one or more annular rings in the power spectrum, thus targeting regions responsible for a given ``powder diffraction'' signature. Alternatively, one might cross-correlate intensities found in patches associated with a specific pair of reflections, thus searching for cross-fringes from a particular single crystal lattice orientation in a large image.

\section{Tiling strategies and decomposition heirarchies}

\begin{figure}
\includegraphics{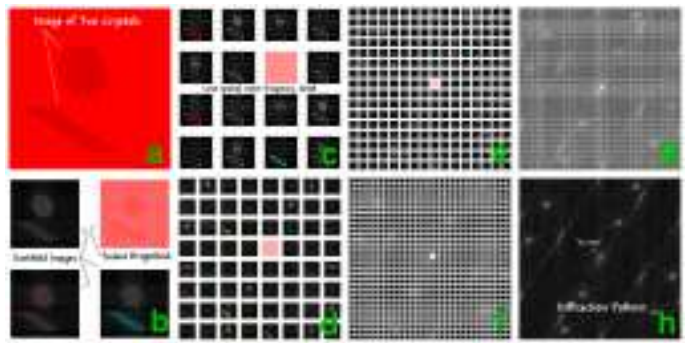} 
\caption{Decomposition series of $n \times n$ square Fourier window decompositions running from a real direct space image in (a) with $n=1$ to the Fourier transform in (h) with $n=128$.  Hue denotes the Fourier phase of each pixel, with red and cyan corresponding to the positive and negative reals, respectively.}
\label{FigSeries1}
\end{figure}

On the subject of decomposition heirarchies (or Daubechies' ``frames'' formed using a discrete sublattice), consider the attached image files in sequence. Figure \ref{Fig2}a illustrates the traditional array of horizontal (01), vertical (10) and diagonal (11) components generated as a 2D image is broken down into wavelets (Axx) along with a rescaled (reduced resolution, or ``brightfield'') image in the upper left corner, which of course may be as shown itself decomposed (Bxx, etc.) at each stage with less direct space but more reciprocal space resolution. One might characterize this as a series of $2 \times 2$ decompositions.

Figure \ref{Fig3}b simply rearranges that $2 \times 2$ heirarchy according to the ``locations'' in frequency space about which each frame is centered, with the DC peak in the field located in the center. Note that the first stage (A) components correspond to higher frequencies, and also higher bandwidths (more resolution) than do the higher stage (e.g. B, C) components -- a feature that distinguishes wavelets from short time Fourier decompositions, for example, and makes them better for wide dynamic range phenomena and ``situations where better time-resolution at high frequencies than at low frequencies is desirable'' \cite{Daubechies90}. Note also, however, that the angular resolution of frequency (subtended at the DC peak) at any given decomposition stage is very poor.

Figure \ref{Fig4}a does the same ``recentering'', except this time for a $4 \times 4$ decomposition heirarchy. Here, the angular resolution is improved, and the ``wraparound'' edge effects (most noticable at the corners) affect fewer elements. Because frequency as well as spatial resolution are crucial in our work, if we have $n \times n$ images where $n{>}16$, then the optimum tiling for us is not $2 \times 2$ or $4 \times 4$, but something like $\sqrt n  \times \sqrt n$. Figures \ref{Fig4}b and \ref{Fig4}c sketch the $8 \times 8$ and $16 \times 16$ decompositions in this series. Remarkably, in a way which may be largely independent of the basis functions involved, when this decomposition is taken to it's $n \times n$ limit on an $n \times n$ image, one gets simply the Fourier transform itself as illustrated in Fig \ref{FigSeries1}.

None of the foregoing discusses the functions used for the decompositions, but only their locations and domains in reciprocal (and scale) space. The simplest conceivable functions to ``center and scale'' to these domains are sharp edged window functions. This is especially true when constructing electron optical filters: You make for example a 10 micron hole in a thin piece of metal, and electrons are either transmitted with unit amplitude and no phase shift where the hole is, or they are not transmitted at all. Partial transparency, or phase modulation, would require controlled aperture thicknesses at the atomic scale, and hence have not historically been a practical choice. The ``darkfield perspective'' on position/frequency localization is of course not limited to such ``all or none'' aperture functions, but they are a natural place to start because of their convenience and history of use in microscopy.

\begin{figure}
\includegraphics{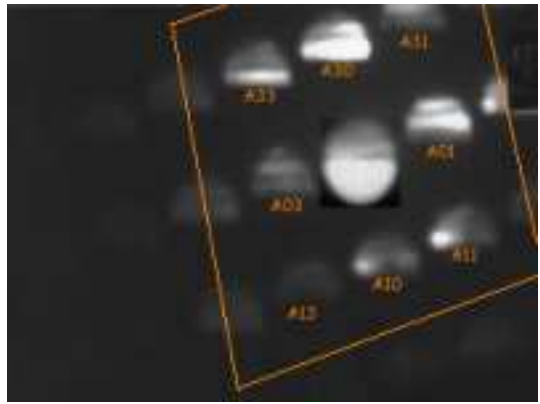} 
\caption{Defocussed electron diffraction pattern of a crystal, showing that sinc-decompositions intermediate between direct and frequency space can be found in nature.}
\label{Fig7}
\end{figure}

\begin{figure}
\includegraphics{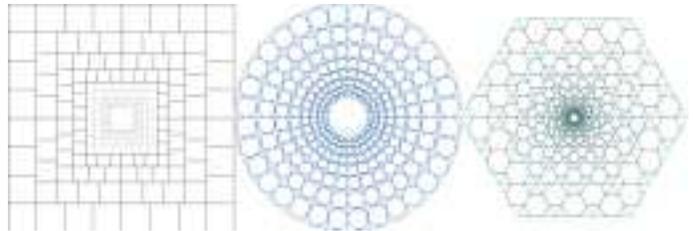} 
\caption{Logarithmic Fourier window tilings with (a) four-fold symmetry in both the initial (direct space) image as well as in the Fourier windows used, (b) polar symmetry Fourier windows applied to an initial circular image source region, and (c) alternating equi-triangular Fourier windows applied to an initial hexagonal image source region.}
\label{Fig911}
\end{figure}

The natural path between direct image and Fourier transform, that this way of looking at hierarchies provides, is more than just theoretical. Figure \ref{Fig7} shows what happens when one forms an image of the wave field in an electron microscope {\em just above or below} the objective back focal plane of a periodic specimen (in this case sapphire) under plane parallel illumination. (The Fourier transform itself resides {\em in} the back focal plane.) As you can see, the decomposition of the previous figures can (it would seem) be found grinning back at us from the midst of the electron wavefield in our scope! Note that here the circular pattern of each inset maps the effective shape of the specimen region being illuminated.  Here as in later examples, symmetries for the direct space source region, and the Fourier tiling, may be chosen independently (cf. Figs. \ref{Fig911}). 

\begin{figure}
\includegraphics{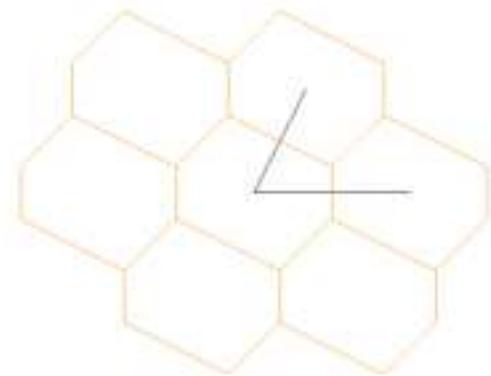} 
\caption{Reciprocal space domains associated with a diffraction spot array from the 2D reciprocal lattice of an arbitrary crystal lattice, viewed in projection with small Bragg angles down a low-symmetry zone axis.}
\label{Fig8}
\end{figure}

Also note that the symmetry of the ``tiling'' is not chosen with apertures, but by the specimen itself. This suggests to me that the wavefield in the microscope may be more generally modeled as a kind of image/FT convolution (is Wigner-Ville transformation relevant here?). It masquerades as a Fourier-space tiling here, because this specimen's ``point-rich'' 3D reciprocal lattice acts (via it's intersection with the incident electron Ewald sphere) as a 2D reciprocal comb (i.e. as a periodic array of delta functions in frequency space).

The ``microscopist's perspective'' also provides some clues to the types of more sophisticated Weyl-Heisenberg, Wigner-Ville and/or wavelet-like decompositions (e.g. Meyer-Coifman brushlets \cite{MeyerCoifman97}) that might prove useful in helping microscopists make the most of their digital images (especially those that show atomic lattice resolution). In this latter case, the fact that the spacing between all atoms in solids is on the order of 0.2 nm means that precise information on both location and scale size may be desired with a very specific frequency range in mind. The structures we search for (discrete arrays of equally spaced molecules) are also very specific. It turns out in this context, for example, that simple rules define a continuum of tilings between square and hexagonal, that have been optimized (as in Fig. \ref{Fig8}) to offer maximum expression of the shape transforms associated with each spot in an arbitrary 2D lattice.  Of course, this strategy is only useful if you know the periodicity and orientation of the lattice of interest.  In the absence of this information, or in the presence of more than one crystal, the symmetry of the tiling scheme may be dictated by other considerations (or the lack thereof).

Also, specific periodicities and their harmonics is a specialized interest. If on the other hand adaptivity to a wide range of scale sizes takes precedence (especially relevant for images greater than a gigabyte in size), logarithmic tilings as in 
Fig \ref{Fig911}, which extend the wavelet strategy of treating frequency as a scale parameter, may be of help.  Many ridgelet tilings, for example, adopt this tact.

In Figure \ref{Fig911}b, the dark circles represent (as do the discs in Figure \ref{Fig7}) the shape in direct space of the region being analyzed, while the light blue patches in which they are centered represent the frequency domains from which these "darkfield image patches" draw power. Likewise for the dark hexagons in Figure \ref{Fig911}c. One of the nice things about sharp window frequency space decompositions (like those commonly used for electron microscope darkfield imaging) is that digital algorithms exist today (some quite efficient) for tiling schemes like those of Figs \ref{Fig911}-\ref{Fig8}.

\section{Digital Darkfield Application Examples}

The examples here illustrate (i) the use of sinc-windowlet intensity maps to highlight the distribution of a selected periodicity in an image, (ii) the use of tiled decompositions to search for patterns of periodicity distribution in space which might otherwise be ``covered'' by noise, and (iii) the use of sinc-windowlet phase maps to plot selected strain components associated with each of several lattice planes whose periodicities are projected into a single image.  All begin with images taken using a 300kV Philips EM430ST transmission electron microscope (TEM) with point resolution near 0.2 nm.

\subsection{Visual tracking of thin-film nucleation and growth}

A look at Cu$_2$O electrochemically grown epitaxially onto the surface of a (100) silicon wafer by Jay Switzer (cf. Switzer et. al. \cite{Switzer99}) at the University of Missouri in Rolla (UMR) illustrates the usefulness of sinc-windowlet analysis, both in identifying barely visible periodicities, and in using digital darkfield images to highlight periodicity distributions.  The specimen was prepared in cross-section by cleaving along (110), followed by preparation of a 3 mm diameter ``stack'' for slicing, abrasion, dimpling, and Argon ion milling.  

\begin{figure}
\includegraphics{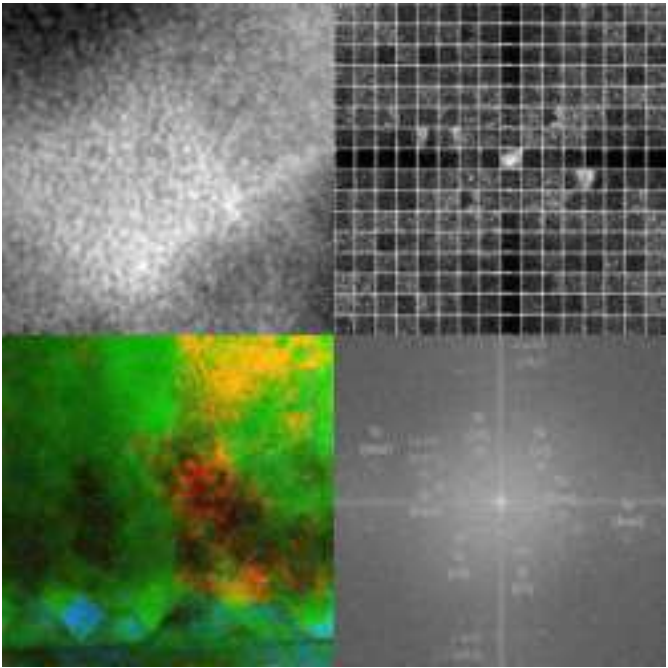} 
\caption{Top left: $512 \times 512$ excerpt from $15000 \times 15000$ image of a Cu$_2$O/Si interface (running from around 7 o'clock to 1 o'clock just to the right of the region shown in the image) in cross-section. This excerpt shows parts of two pyramidal Si mounds on the right, and some fringes on the left from a cuprous oxide column that likely nucleated between them (followed by columnar growth orthogonal to the substrate silicon wafer, viewed here down $\left\langle {110} \right \rangle $);  Right side:  maps calculated from Fig \ref{Fig12}:  On top is a $16 \times 16$ darkfield decomposition showing spatial correlations (left Cu$_2$O, right Si) within selected frequency windows.  At bottom is a power spectrum for measuring periodicity spacings and angles;  Bottom left: Color overlay from a larger partion of the HREM image discussed here, rotated so that the Cu$_2$O/Si interface is horizontal.  The orginal HREM image is in green.  A darkfield version of the same image, formed using only intensity from Si lattice periodicities, is in blue.  In red, find a darkfield image formed using only intensity from the particular 4-degree rotated Cu$_2$O lattice discovered in the other figures}
\label{Fig12}
\end{figure}

An excerpt of a larger high resolution TEM electron phase contrast (HREM) image, covering a region just to the left of the primary Cu$_2$O/Si interface, is shown in Figure \ref{Fig12} top left.  The right side of this image shows parts of two octahedral Si pyramids with (111) faceted walls.  The larger image shows that these pyramids jut upward from the (100) silicon surface, pointing in the direction of Cu$_2$O epitaxy.  The left side of the image, by comparison, contains lattice fringes from the Cu$_2$O layer itself.  Of particular interest is the question of growth nucleation.  These two phases have a large lattice misfit, and the Cu$_2$O itself shows evidence of low angle grain boundaries in the epitaxy, suggesting that a mechanism for accomodating misfit through disorder is at work.  

Unfortunately, the presence of amorphous material (e.g. silicon oxide) as well as a 50 nm layer on top and bottom of the specimen due to ion milling are conspiring to make the lattice fringes fairly difficult to see or track throughout the image.  To get a closer look, therefore, we calculated a $16 \times 16$ darkfield decomposition of this image (Fig \ref{Fig12} top right) along with a regular Fourier power spectrum (Fig \ref{Fig12} bottom right).  From the power spectrum, of course, interspot spacings and angles allows us to compare fringe spacings and angles rather quantitatively.  The darkfield decomposition, by comparison, allows us to tell where in the image the "power" or variance, in a given patch of frequency space, is concentrated.  


The eight blue circles in the top right of Figure \ref{Fig12} highlight regions of frequency space whose variations are concentrated in the right half of the image, i.e. in the silicon pyramids.  One can tell this because in these Fourier windows the right half of the patch is noticably brighter than the left half.  The four orange circles highlight regions of frequency space whose variations are concentrated in the left half of the image, i.e. in the Cu$_2$O lattice.


This is confirmed by a look at these corresponding regions of frequency space in the power spectrum in Fig \ref{Fig12} (bottom right).  Measurement of the spacings and angles confirms that the blue-labeled spots correspond to Si spacings viewed down the $\left\langle {011} \right \rangle $ zone, with the (400) reflection perpendicular to the Cu$_2$O/Si interface in projection.  The orange spots correspond to a Cu$_2$O $\left\langle {011} \right \rangle $ zone, with the (200) reflection about 4 degrees clockwise from the interface normal.  The Cu$_2$O lattice is also twisted between 0.5 and 1 degree ``up'' about the beam normal, because the (020) spot is not visible while the (040) spot is there if extremely weak.  In fact, this 1.07 Angstrom (040) spot would not have been noticed or confirmed without the evidence for spatial correlation with the Cu$_2$O side of the image in the corresponding frequency window of the darkfield decomposition at top.  Given this spot, we are able to determine the orientation of the Cu$_2$O column to better than a degree {\em in any direction}.  Future work with even more careful filtering might allow inspection of the epitaxy between Cu$_2$O and Si, for evidence of dislocations at points of Cu$_2$O/Si contact.

Work on the larger image from which this was drawn confirmed that the ``4 degree off'' column of Cu$_2$O indeed extends up to the top of the layer.  But where near the interface does it begin, and how can we map the location of this particular crystal ``from a distance'', i.e. in a figure sufficiently reduced in size that the whole film can be seen even though the lattice fringes themselves are then too small to resolve?

Figure \ref{Fig12} bottom left illustrates with a composite which began with the reduced image of a larger part of the original negative in green.  Overlaid in blue find a darkfield image created by adding intensities from Fourier windows which pass primarily periodicities of the silicon substrate.  As you can see, the silicon pyramids with thin silicon at the top show up especially well.  Lastly, in red (which combines with green to form a coppery orange) find a darkfield image created by adding intensities from Fourier windows which pass primarily periodicities of the ``4 degree off'' Cu$_2$O column.  The image confirms the columnar growth pattern, and moreover offers some reason to suggest that the epitaxy probably takes place near the peak of the Si pyramids, rather than at their base.  This would offer opportunities for the Cu$_2$O lattice to attach near the tip of this pyramid (or along it's side), taking advantage of the high local curvature to select a minor tilt to help accomodate the lattice misfit.  In this way, digital darkfield imaging might allow one to visualize processes on scales much larger than the periodicities being effected.  We expect that this use will become more important as even larger images, with atom-scale metric information in them, become available.

\subsection{Single image recognition of icosahedral twins}

\begin{figure}
\includegraphics{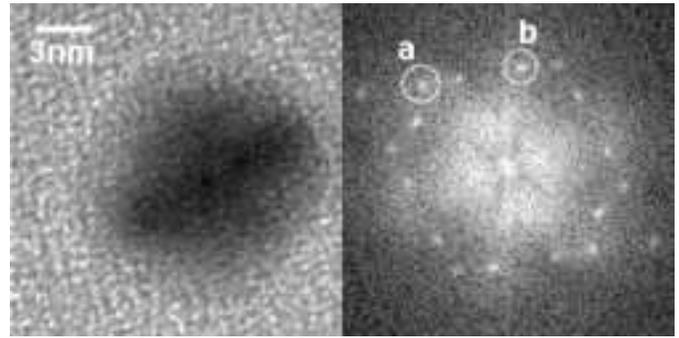}
\caption{Left: HREM image of a Pd icosahedral twin formed by solid state reduction in polyvinyl chloride; Right: Power spectrum of the above image showing elements of 10-fold symmetry.}
\label{Fig_i1}
\end{figure}

\begin{figure}
\includegraphics{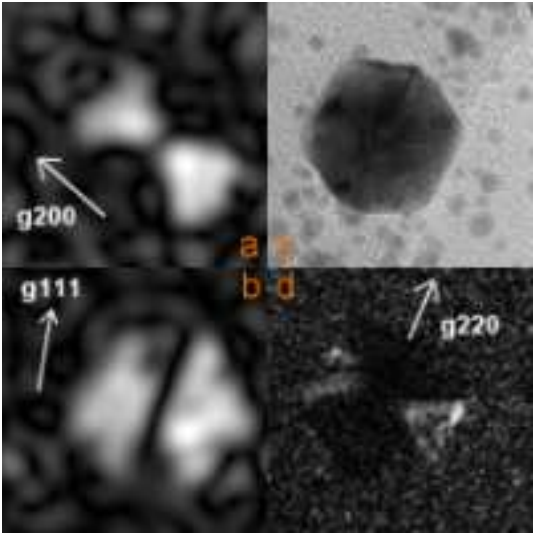}
\caption{a: Digital darkfield image using a Pd (200) reflection, showing the predicted ``bowtie'' twin pair;  b: Digital darkfield image using a Pd (111) reflection, showing the predicted ``butterfly'' twin quartet;  c: $1024 \times 1024$ HREM image of a ``randomly oriented'' and twinned AuPd cluster;  d: Darkfield 0.14 nm (220) butterfly, discovered in a $32 \times 32$ darkfield search of the image above for i-twin ``bowties'' and ``butterflies''.}
\label{Fig_i2}
\end{figure}

In this section, we treat a more complex problem, namely the study of twinned nanoparticles.  Although the stable configuration for large crystals of a given metal at room temperature is often well known, when metal particles are smaller than 10 nm in size, surface to volume ratios are much larger and modes of crystal growth not energetically favorable (or even impossible) {\em in bulk} are quite common.  The classic example of this is icosahedral growth, perhaps facilitated by the low surface area of triangular fcc (111) facets in an icosahedral array \cite{Hofmeister98}.  Surface considerations begin to pale as such particles acquire volume, and what began as an icosahedral assemblage grows into a set of 20 paired and nearly tetrahedral fcc twins (one twin for each icosahedral face).

The HREM image of a particle of icosahedrally twinned fcc Palladium, formed by solid state reduction in polyvinyl chloride, microtomed, and viewed down the 5-fold axis, is shown in the top of Fig \ref{Fig_i1}.  Although symmetry in the structure may not be apparent from the image, the central region of the power spectrum (bottom Fig \ref{Fig_i1}) reveals a clear 10-fold, or pseudo 10-fold, symmetry.  The inner-most spacings are consistent with 0.22 nm fcc Pd (111), the intermediate spots are consistent Pd (200) spacings of around 0.19 nm, and the occasional outer spot is consistent with the 0.14 nm Pd (220).  Of course, fcc crystals (in fact, all translationally periodic lattices in 3D) have no 5 or 10 fold orientations.


Digital darkfield images, formed using the periodicities surrounding a single spot in the Fourier transform, provide a striking answer to the question:  What's going on?  For example, the top left of Figure \ref{Fig_i2} was obtained by windowing the region of frequency space marked (a) in Fig \ref{Fig_i1}.  As a result, the set of Pd (220) planes running from lower left to upper right in one pair of twins in the particle ``lights up''.  Electron microscopists might see this as ``forming the image with electron Bragg scattered into the Fourier aperture by those two pyramidal crystals'', although of course no diffraction is taking place in this analysis at all.  Similarly, each set of (111) spacings in an ideal fcc icosahedral twin is shared by four of these crystals.  When viewed down a 5-fold direction, such (111) planes ``light up'' a butterfly pattern like that in the bottom left of Fig \ref{Fig_i2}.  This image was obtained by windowing only those frequencies surround the spot marked (b) in the power spectrum of Fig \ref{Fig_i1}.

The images in Figure \ref{Fig_i2} will be familiar to materials electron microscopists as darkfield images with ``the g-vector of the active reflection'' drawn in (cf. Hirsch et al. \cite{Hirs11}.  The small size of this twin particle, and the crowded nature of it's reciprocal lattice, would make this experiment difficult to perform optically, but the strategy is familiar.  Consider now, however, the problem of identifying icosahedral twinning in a randomly-oriented nano-particle.  Individual 2-second exposure HREM images can include lattice information on hundreds of nanoparticles, but can harmonic analysis allow one to detect icosahedral twinning after the fact?

The top right of Figure \ref{Fig_i2} is a HREM image of a twinned AuPd cluster provided by Massimo Bertino at UMR (cf. Doudna et al. \cite{Doudna01}).  The specimen is on a carbon support film.  Its power spectrum (not shown) contains spot spacings characteristic of AuPd, but with apparently random orientation thus offering no evidence of symmetry in the specimen.  However, a $32 \times 32$ tiled darkfield decomposition of this $1024 \times 1024$ image (analogous to that in Fig \ref{Fig12}) contained at least one bowtie/butterfly structure like those in Fig \ref{Fig_i2}.  The clearest of these (a bowtie) is shown in the bottom right of \ref{Fig_i2}, where the g-vector orientation associated therewith is the one expected for icosahedral twinning as well.  

The numerical calculation of 1023 tiled darkfield images (and one brightfield image) which made this discovery possible was done in essentially the time required for a single forward and reverse $1024 \times 1024$ fast Fourier transform.  We are currently doing calculations to determine the fraction of such icosahedral twins likely to be indentifiable from a single, randomly oriented, image.  Back of the envelope considerations, as well as the data shown here, suggest that the method has a signficant chance of success.  

Even if one could optically detect diffraction spots from this particle, and set up darkfields for each of the diffraction spots in the pattern, the time involved would be prohibitive.  Since this calculation can also be done as well for the hundred or more other particles represented in the quarter gigabyte of information from a single 2-second HREM image exposure, this strategy of tiled darkfield decomposition shows potential for offering both extended sensitivity, and major savings of time, in the study of nanoparticle assemblages.

\subsection{Mapping lattice strain around 2D and 3D defects}

\begin{figure}
\includegraphics{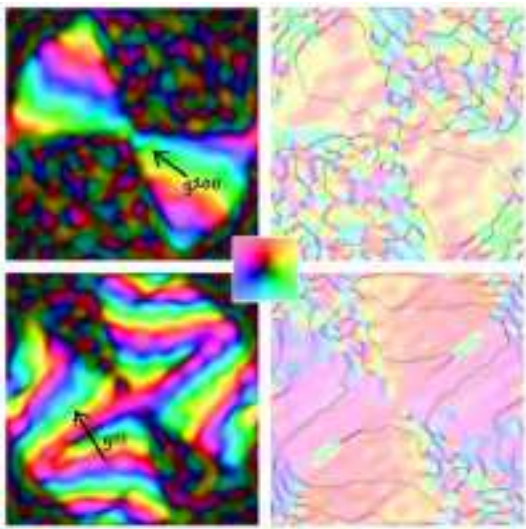}
\caption{Complex color phase (left) and strain (right) maps for bowties (top) and butterflies (bottom) from a simulated icosahedral twin.}
\label{Fig_i4}
\end{figure}

Thanks in part to the fact that transmission electron microscopes record data ``in parallel'' as distinct from ``in series'' (as is the case e.g. for scanning probe microscopes), HREM images under optimum conditions can map the projected potential in a thin specimen with reasonably high metric fidelity.  Hence distortions measured in HREM lattice images can provide rather precise information on lattice structure in a specimen \cite{Seitz98}.  In this section, we discuss how the Fourier phase of darkfield images can therefore provide details about specimen structure e.g. on percent-level lattice strain components as projected into the beam direction.  The extent to which the process of thinning the specimen, as well as strain relaxation at the surface after thinning, modifies the state of the observed specimen from it's state e.g. prior to thinning \cite{Treacy86, Gibson87} will not be discussed.  We will also here ignore effects, like local specimen thickness or instrument defocus, which might in themselves result in shifts in the image Fourier phase from point to point.  Thus we will content ourselves with the mathematician's task of localizing periodicity in the images, and leave to others the task of inferring what those periodicities might mean for the specimen itself.

Let's begin with a theoretical example.  The left half of Figure \ref{Fig_i4} contains maps for bowtie (top) and butterfly (bottom) darkfield images from a modeled icosahedral twin particle, rendered in ``logarithmic complex color''.  This means that brightness in the image is logarithmically related to the windowlet amplitude, while the phase is rendered using a cyclic hue-related color table (see inset) in which red (for example) might correspond to positive real and cyan negative real.  A similar color scheme was earlier used in Figure \ref{FigSeries1} to represent complex numbers.  

Fourier phase shifts in a periodicity amount to local displacements.  For example, in one spatial dimension an interval with sinusoidal oscillations would be shifted by $\pi /2$ from an adjacent interval with cosine-usoidal variation.  If, as in the cases discussed here the darkfield image is dominated by power in a single lattice periodicity, then the 2D gradient in that phase describes the 2D strain one-form, the components of which are simply change in lattice position per unit traverse in a given direction.  The log magnitude of this strain one-form is inversely proportional to brightness in the images on the right side of Fig \ref{Fig_i4}, while the direction of that strain is denoted by hue (as per the key in the center of the figure).

\begin{figure}
\includegraphics{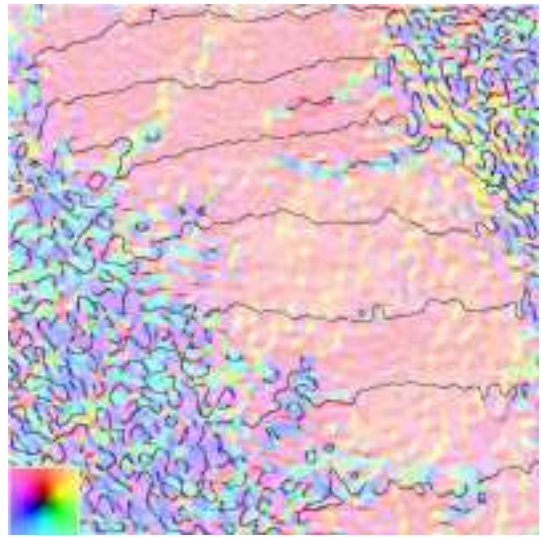}
\caption{Log reciprocal-amplitude directional complex-color strain image for the GaN (003) ``growth direction'' periodicity which runs perpendicular to the GaN/sapphire interface (upper left corner).}
\label{Fig_i5}
\end{figure}

Note that the strain seems random in areas with windowlet amplitudes down in the noise (i.e. away from the object with strong periodicities in this frequency window).  In these areas, strain fluctuations occur on a scale size inversely proportional to the bandwidth of the darkfield window.  In regions of high amplitude the strain by comparison seems relatively uniform, i.e. on the bow-tie and butterfly patterns themselves.  However, note the blue bands centered on the butterfly wings in the lower right quadrant of Fig \ref{Fig_i4}.  This is the strain associated with the tensional (outer) half of the partly compressional, partly tensional (111) interface between adjacent ``bowties'' (paired crystal 3-sided pyramids).  It arises in essence because the fcc geometry is asking for a $70.5^o$ ($\leftarrow \arccos \left[ 1/3 \right]$) angle between similar sections, while the icoshedral symmetry demands $72^o$ ($\leftarrow 2 \pi /5$).  Note that the direction (hence hue) of the interface strain shows it to be directed perpendicular to the interface between crystals (as well as the planes associated with the windowed periodicity).  The short range of the strain is a result of the detailed icosahedral model used: No ``relaxation length'' was included in the calculation.

This theoretical example in hand, consider now an experimental example.  Figure \ref{Fig_i5} shows the strain image obtained as described above, for the (003) growth direction periodicity at the bottom of a ``GaN/InGaN on sapphire'' molecular beam epitaxy quantum well specimen (cf. Nistor et al. \cite{Nistor00}) grown and provided by Dan Leopold.  Both the sapphire in the upper right corner, and the ion-mill damaged thin area to the lower left, are noise dominated in the chosen frequence window.  The rest of the image, however, is quite uniform in color save for the dark near-horizontal lines across the specimen.  These represent artificial ``branch cuts'' in the phase values which gave rise (unnecessarily) to singularities during the gradient calculation.  The magnitude of the strain in this GaN (003) layer is at the percent level (i.e. measurable in picometers for each 0.27 nm period of traverse in the image), and the actual value of lattice strain in the growth direction is probably less given that a shift in the reference periodicity would likely reduce the coloration even further.  

The same starting image can also be used to examine strain components in other periodicities.  For example, strains in the GaN (110) direction (parallel to the interface in this projection) are comparable but also a bit noisier, because the microscope transfer function deliver's the 0.17 nm periodicity less efficiently than the 0.27 nm periodicity in the growth direction.  

The strategy described here, sensitive to ``sub-pixel displacements'' like the one recommended by Seitz et al. \cite{Seitz98}, thus arises naturally in the context of a comprehensive darkfield analysis.  Problems with overlapping periodicities, noise, and microscope instability nonetheless continue to limit our inferences.  Hence these applications also provide a context for new harmonic analysis strategies that will ``divide up reciprocal space'' in ways which are more informed, to the structures being examined, than the ``sharp-edged window method'' applied here.  We know in advance that none of the features sought in microscope images are likely to have sharp-edged window footprints in frequency space.

\section{Conclusions and Challenges}

  This paper is designed primarily for two audiences.  For those involved in mathematical harmonic analysis, it is designed to illustrate a few applications of interest in microscopy on the nanoscale, for which optical sinc-windowlet analyses are in a relatively advanced stage, but for which digital solutions are in their infancy.  It is also designed to highlight the windowlet interpretation of intermediate direct and frequency space decompositions.  In this interpretation, {\em all} ``energy conserving'' decompositions are in a physical sense dividing up the frequency space ``wavefield'' of the orginal dataset, thus providing the analyst with ``darkfield images'' chosen to highlight the image patterns of interest.  For those involved in microscopy, the paper is designed to illustrate digital versions of familar optical darkfield analysis techniques which, as in electron optic applications in principle, use ``sharp-edged'' apertures.

  Challenges here posed for harmonic analysis include (i) the application of existing frequency-selective decompositions (e.g. ridgelets and brushlets) to the study of nanocrystals as well as nanocrystal edges and other defects, (ii) the possible use of radon, edgelet, and other decompositions for the recognition of single-walled structures (e.g. graphene sheets) in a samples ranging from carbon composites to dust formed in the outer atmosphere of red giants, and (iii) future development of decompositions optimized for taking advantage of phase and amplitude information in HREM images to recognize bump (single atom) lattices buried in noise, and perhaps even order in paracrystalline \cite{Gibson3} or even polymerized solids.  This opportunity, to drive discovery between active but largely independent communities of investigators, will hopefully benefit other application areas as well.

\begin{acknowledgments}
Thanks to Max Bertino for the AuPd clusters, Jay Switzer for the copper
oxide on silicon epitaxy, and Dan Leopold for the GaN on sapphire 
specimens.  Thanks to Grant Welland, Shuhan Lin, and others across 
the region for insights and shared enthusiasm for the challenge of detective 
work on small worlds.  This work has been supported in part by instrumentation 
and funding from Monsanto and MEMC Electronic Materials Companies.
\end{acknowledgments}

\bibliographystyle{apsrev}
\bibliography{harmonic.bib}

\begin{thebibliography}{20}
\expandafter\ifx\csname natexlab\endcsname\relax\def\natexlab#1{#1}\fi
\expandafter\ifx\csname bibnamefont\endcsname\relax
  \def\bibnamefont#1{#1}\fi
\expandafter\ifx\csname bibfnamefont\endcsname\relax
  \def\bibfnamefont#1{#1}\fi
\expandafter\ifx\csname citenamefont\endcsname\relax
  \def\citenamefont#1{#1}\fi
\expandafter\ifx\csname url\endcsname\relax
  \def\url#1{\texttt{#1}}\fi
\expandafter\ifx\csname urlprefix\endcsname\relax\def\urlprefix{URL }\fi
\providecommand{\bibinfo}[2]{#2}
\providecommand{\eprint}[2][]{\url{#2}}

\bibitem[{\citenamefont{Goupillaud et~al.}(1984)\citenamefont{Goupillaud,
  Grossmann, and Morlet}}]{Goupillaud84}
\bibinfo{author}{\bibfnamefont{P.}~\bibnamefont{Goupillaud}},
  \bibinfo{author}{\bibfnamefont{A.}~\bibnamefont{Grossmann}},
  \bibnamefont{and} \bibinfo{author}{\bibfnamefont{J.}~\bibnamefont{Morlet}},
  \bibinfo{journal}{Geoexploration} \textbf{\bibinfo{volume}{23}},
  \bibinfo{pages}{85} (\bibinfo{year}{1984}).

\bibitem[{\citenamefont{Hirsch et~al.}(1977)\citenamefont{Hirsch, Howie,
  Nicholson, Pashley, and Whelan}}]{Hirs11}
\bibinfo{author}{\bibfnamefont{P.}~\bibnamefont{Hirsch}},
  \bibinfo{author}{\bibfnamefont{A.}~\bibnamefont{Howie}},
  \bibinfo{author}{\bibfnamefont{R.~B.} \bibnamefont{Nicholson}},
  \bibinfo{author}{\bibfnamefont{D.~W.} \bibnamefont{Pashley}},
  \bibnamefont{and} \bibinfo{author}{\bibfnamefont{M.~J.}
  \bibnamefont{Whelan}}, \emph{\bibinfo{title}{Electron Microscopy of Thin
  Crystals}} (\bibinfo{publisher}{Robert E. Krieger Publishing Company},
  \bibinfo{address}{Huntington, New York}, \bibinfo{year}{1977}), pp.
  \bibinfo{pages}{310--311}.

\bibitem[{\citenamefont{Daubechies}(1990)}]{Daubechies90}
\bibinfo{author}{\bibfnamefont{I.}~\bibnamefont{Daubechies}},
  \bibinfo{journal}{IEEE Trans. Inf. Theor.}
  \textbf{\bibinfo{volume}{36}}(\bibinfo{number}{5}), \bibinfo{pages}{961}
  (\bibinfo{year}{1990}).

\bibitem[{\citenamefont{Chui}(1996)}]{Chui96}
\bibinfo{author}{\bibfnamefont{C.~K.} \bibnamefont{Chui}},
  \bibinfo{journal}{Bull. Amer. Math. Soc. (N.S.)}
  \textbf{\bibinfo{volume}{33}}(\bibinfo{number}{1}), \bibinfo{pages}{131}
  (\bibinfo{year}{1996}).

\bibitem[{\citenamefont{Boone}(1998)}]{Boone98}
\bibinfo{author}{\bibfnamefont{B.~G.} \bibnamefont{Boone}},
  \emph{\bibinfo{title}{Signal Processing in Optics}}
  (\bibinfo{publisher}{Oxford University Press}, \bibinfo{address}{New York},
  \bibinfo{year}{1998}).

\bibitem[{\citenamefont{Flesia et~al.}(2001)\citenamefont{Flesia, Hel-Or,
  Averbuch, Candes, Coifman, and Donoho}}]{Flesia01}
\bibinfo{author}{\bibfnamefont{A.~G.} \bibnamefont{Flesia}},
  \bibinfo{author}{\bibfnamefont{H.}~\bibnamefont{Hel-Or}},
  \bibinfo{author}{\bibfnamefont{A.}~\bibnamefont{Averbuch}},
  \bibinfo{author}{\bibfnamefont{E.~J.} \bibnamefont{Candes}},
  \bibinfo{author}{\bibfnamefont{R.~R.} \bibnamefont{Coifman}},
  \bibnamefont{and} \bibinfo{author}{\bibfnamefont{D.~L.}
  \bibnamefont{Donoho}}, \emph{\bibinfo{title}{Digital implementation of
  ridgelet packets}} (\bibinfo{publisher}{Academic Press},
  \bibinfo{address}{New York}, \bibinfo{year}{2001}).

\bibitem[{\citenamefont{Vetterli and Kovacevic}(1995)}]{Vetterli95}
\bibinfo{author}{\bibfnamefont{M.}~\bibnamefont{Vetterli}} \bibnamefont{and}
  \bibinfo{author}{\bibfnamefont{J.}~\bibnamefont{Kovacevic}},
  \emph{\bibinfo{title}{Wavelets and Subband Coding}}
  (\bibinfo{publisher}{Prentice Hall}, \bibinfo{year}{1995}).

\bibitem[{\citenamefont{Calderbank et~al.}(1998)\citenamefont{Calderbank,
  Daubechies, Sweldens, and Yeo}}]{Daubechies98}
\bibinfo{author}{\bibfnamefont{R.}~\bibnamefont{Calderbank}},
  \bibinfo{author}{\bibfnamefont{I.}~\bibnamefont{Daubechies}},
  \bibinfo{author}{\bibfnamefont{W.}~\bibnamefont{Sweldens}}, \bibnamefont{and}
  \bibinfo{author}{\bibfnamefont{B.-L.} \bibnamefont{Yeo}},
  \bibinfo{journal}{Applied and Computational Harmonic Analysis (ACHA)}
  \textbf{\bibinfo{volume}{5}}(\bibinfo{number}{3}), \bibinfo{pages}{332}
  (\bibinfo{year}{1998}),
  \urlprefix\url{citeseer.nj.nec.com/article/calderbank96wavelet.html}.

\bibitem[{\citenamefont{Meyer and Coifman}(1997)}]{MeyerCoifman97}
\bibinfo{author}{\bibfnamefont{F.~G.} \bibnamefont{Meyer}} \bibnamefont{and}
  \bibinfo{author}{\bibfnamefont{R.~R.} \bibnamefont{Coifman}},
  \bibinfo{journal}{Applied and Computational Harmonic Analysis}
  \textbf{\bibinfo{volume}{4}}, \bibinfo{pages}{147} (\bibinfo{year}{1997}).

\bibitem[{\citenamefont{Fraundorf and Pollack}(1991)}]{Fraundorf91}
\bibinfo{author}{\bibfnamefont{P.}~\bibnamefont{Fraundorf}} \bibnamefont{and}
  \bibinfo{author}{\bibfnamefont{K.}~\bibnamefont{Pollack}},
  \bibinfo{journal}{Ultramicroscopy} \textbf{\bibinfo{volume}{37}},
  \bibinfo{pages}{72} (\bibinfo{year}{1991}).

\bibitem[{\citenamefont{Marks}(1996)}]{Marks96}
\bibinfo{author}{\bibfnamefont{L.~D.} \bibnamefont{Marks}},
  \bibinfo{journal}{Ultramicroscopy} \textbf{\bibinfo{volume}{62}},
  \bibinfo{pages}{43} (\bibinfo{year}{1996}).

\bibitem[{\citenamefont{Fraundorf}(1990)}]{Fraundorf90}
\bibinfo{author}{\bibfnamefont{P.}~\bibnamefont{Fraundorf}},
  \bibinfo{journal}{Phys. Rev. Lett.}
  \textbf{\bibinfo{volume}{64}}(\bibinfo{number}{9}), \bibinfo{pages}{1031}
  (\bibinfo{year}{1990}).

\bibitem[{\citenamefont{Switzer et~al.}(1999)\citenamefont{Switzer, Shumsky,
  and Bohannan}}]{Switzer99}
\bibinfo{author}{\bibfnamefont{J.~A.} \bibnamefont{Switzer}},
  \bibinfo{author}{\bibfnamefont{M.~G.} \bibnamefont{Shumsky}},
  \bibnamefont{and} \bibinfo{author}{\bibfnamefont{E.~W.}
  \bibnamefont{Bohannan}}, \bibinfo{journal}{Science}
  \textbf{\bibinfo{volume}{284}}(\bibinfo{number}{5412}), \bibinfo{pages}{293}
  (\bibinfo{year}{1999}).

\bibitem[{\citenamefont{Hofmeister}(1998)}]{Hofmeister98}
\bibinfo{author}{\bibfnamefont{H.}~\bibnamefont{Hofmeister}},
  \bibinfo{journal}{Cryst. Res. Technol.}
  \textbf{\bibinfo{volume}{33}}(\bibinfo{number}{1}), \bibinfo{pages}{3}
  (\bibinfo{year}{1998}).

\bibitem[{\citenamefont{Doudna et~al.}(2001)\citenamefont{Doudna, Hund, and
  Bertino}}]{Doudna01}
\bibinfo{author}{\bibfnamefont{C.~A.} \bibnamefont{Doudna}},
  \bibinfo{author}{\bibfnamefont{J.~F.} \bibnamefont{Hund}}, \bibnamefont{and}
  \bibinfo{author}{\bibfnamefont{M.~F.} \bibnamefont{Bertino}},
  \bibinfo{journal}{Intern. J. Mod. Phys. B}
  \textbf{\bibinfo{volume}{15}}(\bibinfo{number}{24-25}), \bibinfo{pages}{3302}
  (\bibinfo{year}{2001}).

\bibitem[{\citenamefont{Seitz et~al.}(1998)\citenamefont{Seitz, Ahlborn, Seibt,
  and Schroter}}]{Seitz98}
\bibinfo{author}{\bibfnamefont{H.}~\bibnamefont{Seitz}},
  \bibinfo{author}{\bibfnamefont{K.}~\bibnamefont{Ahlborn}},
  \bibinfo{author}{\bibfnamefont{M.}~\bibnamefont{Seibt}}, \bibnamefont{and}
  \bibinfo{author}{\bibfnamefont{W.}~\bibnamefont{Schroter}},
  \bibinfo{journal}{J. Microscopy} \textbf{\bibinfo{volume}{190}},
  \bibinfo{pages}{184} (\bibinfo{year}{1998}).

\bibitem[{\citenamefont{Treacy and Gibson}(1986)}]{Treacy86}
\bibinfo{author}{\bibfnamefont{M.~M.~J.} \bibnamefont{Treacy}}
  \bibnamefont{and} \bibinfo{author}{\bibfnamefont{J.~M.}
  \bibnamefont{Gibson}}, \bibinfo{journal}{J. Vac. Sci. Technol. B}
  \textbf{\bibinfo{volume}{4}}, \bibinfo{pages}{1458} (\bibinfo{year}{1986}).

\bibitem[{\citenamefont{Gibson and McDonald}(1987)}]{Gibson87}
\bibinfo{author}{\bibfnamefont{J.~M.} \bibnamefont{Gibson}} \bibnamefont{and}
  \bibinfo{author}{\bibfnamefont{M.~L.} \bibnamefont{McDonald}},
  \bibinfo{journal}{Mat. Res. Soc. Proc.} \textbf{\bibinfo{volume}{82}},
  \bibinfo{pages}{109} (\bibinfo{year}{1987}).

\bibitem[{\citenamefont{Nistor et~al.}(2000)\citenamefont{Nistor, Bender,
  Vantomme, Wu, Landuyt, O'Donnell, Martin, Jacobs, and Moerman}}]{Nistor00}
\bibinfo{author}{\bibfnamefont{L.}~\bibnamefont{Nistor}},
  \bibinfo{author}{\bibfnamefont{H.}~\bibnamefont{Bender}},
  \bibinfo{author}{\bibfnamefont{A.}~\bibnamefont{Vantomme}},
  \bibinfo{author}{\bibfnamefont{M.~F.} \bibnamefont{Wu}},
  \bibinfo{author}{\bibfnamefont{J.~V.} \bibnamefont{Landuyt}},
  \bibinfo{author}{\bibfnamefont{K.~P.} \bibnamefont{O'Donnell}},
  \bibinfo{author}{\bibfnamefont{R.}~\bibnamefont{Martin}},
  \bibinfo{author}{\bibfnamefont{K.}~\bibnamefont{Jacobs}}, \bibnamefont{and}
  \bibinfo{author}{\bibfnamefont{I.}~\bibnamefont{Moerman}},
  \bibinfo{journal}{Appl. Phys. Lett.} \textbf{\bibinfo{volume}{77}},
  \bibinfo{pages}{77} (\bibinfo{year}{2000}).

\bibitem[{\citenamefont{Gibson and Treacy}(1998)}]{Gibson3}
\bibinfo{author}{\bibfnamefont{J.~M.} \bibnamefont{Gibson}} \bibnamefont{and}
  \bibinfo{author}{\bibfnamefont{M.~M.~J.} \bibnamefont{Treacy}},
  \bibinfo{journal}{J. Non-Cryst. Solids} \textbf{\bibinfo{volume}{231}},
  \bibinfo{pages}{99} (\bibinfo{year}{1998}).

\end{thebibliography}

\end{document}